\documentclass{aastex63}

\usepackage{amsmath}
\usepackage{amssymb}
\usepackage{graphicx}

\begin{document}


\title{Basic Pattern of Three-dimensional Magnetic Reconnection within Strongly Turbulent Current Sheets}

\correspondingauthor{Xin Cheng}
\email{xincheng@nju.edu.cn}

\author[0000-0001-9863-5917]{Yulei Wang}
\affiliation{School of Astronomy and Space Science, Nanjing University, Nanjing 210023, People's Republic of China}
\affiliation{Key Laboratory for Modern Astronomy and Astrophysics (Nanjing University), Ministry of Education, Nanjing 210023, People's Republic of China}

\author[0000-0003-2837-7136]{Xin Cheng}
\affiliation{School of Astronomy and Space Science, Nanjing University, Nanjing 210023, People's Republic of China}
\affiliation{Key Laboratory for Modern Astronomy and Astrophysics (Nanjing University), Ministry of Education, Nanjing 210023, People's Republic of China}

\author[0000-0002-4978-4972]{Mingde Ding}
\affiliation{School of Astronomy and Space Science, Nanjing University, Nanjing 210023, People's Republic of China}
\affiliation{Key Laboratory for Modern Astronomy and Astrophysics (Nanjing University), Ministry of Education, Nanjing 210023, People's Republic of China}

\begin{abstract}
Magnetic reconnection is a fundamental mechanism of driving eruptive phenomena of different scales and may be coupled with turbulence as suggested by recent remote-sensing and in-situ observations.
However, the specific physics behind the complex three-dimensional (3D) turbulent reconnection remains mysterious.
Here, we develop a novel methodology to identify and analyze multitudes of multi-scale reconnection fragments within a strongly turbulent current sheet (CS) and apply it to a state-of-the-art numerical simulation of turbulent reconnection for solar flares.
It is determined that the reconnection fragments tend to appear as quasi-2D sheets forming along local magnetic flux surfaces, and, due to strong turbulence, their reconnection flow velocities and reconnection rates are significantly broadened statistically but are scale-independent.
Each reconnection fragment is found to be surrounded by strongly fluctuated in/out-flows and has a widely distributed reconnection rate, mainly in the range of $0.01$--$0.1$.
The results, for the first time, provide quantitative measurements of 3D magnetic reconnection in strongly turbulent flare CSs, offering insights into the cascading laws of 3D reconnection in other turbulent plasmas.
\end{abstract}


\section{Introduction}

Magnetic reconnection occurring in current sheets (CSs) is a fundamental plasma process to drive eruptive phenomena across the universe, during which magnetic energy is rapidly converted to expel fast bulk flows, heat plasmas, and accelerate charged particles \citep{Masuda1994,Matsumoto2015,Shukla2020,Yang2024a,Huang2024,Wang2022e,Ping2023}.
Early two-dimensional (2D) reconnection models, including the Sweet-Parker and Petschek models \citep{Parker1957,Sweet1958,Petschek1964}, assume a stationary and laminar CS.
However, numerous studies over recent decades have highlighted the importance of three-dimensional (3D) effects, revealing that CSs can become highly fragmented and dynamic due to various instabilities, such as the tearing-mode instability (TMI) and Kelvin-Helmholtz instability (KHI), especially in astrophysical plasmas with large magnetic Reynolds numbers \citep{Loureiro2007,Bhattacharjee2009,Daughton2011,Loureiro2013,Huang2016,Kowal2020,Wang2023a}.
On the other hand, magnetic reconnection can be inevitably coupled with turbulence \citep{Lazarian1999}, which not only increases the CS widths but also enhances the reconnection rate \citep{Kowal2009,Kowal2017,Yang2020}.

Solar flares, the most energetic eruptive phenomena in the solar system, have recently been observed to exhibit numerous turbulent features.
Fragmented and dynamic reconnection structures have been identified within the flare CSs that connect coronal mass ejections (CMEs) and flare loops \citep{Warren2018,Cheng2018}.
These structures are likely the fundamental drivers of turbulent flows at the tops of flare loops \citep{Kontar2011,McKenzie2013} and the fine structures observed in flare ribbons \citep{French2021,Wyper2021,CorchadoAlbelo2024,ThoenFaber2025}.
The specific physics of multi-scale 3D reconnection regions constituting turbulent flare CSs, serving as the ``Rosetta Stone'' of understanding specific energy release mechanisms and interpreting observed fine structures, unfortunately, remains poorly understood.

With the great advances in high-performance computers, high-resolution simulations have become an indispensable tool for investigating 3D turbulent reconnection \citep{Ji2022}.
However, accurately identifying and analyzing reconnection regions within a turbulent CS from vast simulation data presents a new challenge, hindering a comprehensive understanding of 3D turbulent reconnection.
In weak magnetic turbulence, where the turbulent magnetic field $\delta B$ is much smaller than the background field $B_0$, the problem can be simplified using a 2D approximation on planes perpendicular to the background field \citep{Zhdankin2013,Li2021b,Li2023b,Dong2022}.
In contrast, in strongly turbulent plasmas where $\delta B \geq B_0$, as sophisticated methods are absent \citep{Vlahos2023,Isliker2019,Kowal2020,Lapenta2021}, determining the nature of 3D reconnection regions, including their geometric shapes, reconnection flows, and reconnection rates, is still in the early stage.
To tackle this problem, we propose a novel methodology to quantitatively analyze 3D turbulent reconnection.
We apply it to the simulation data of the strongly turbulent 3D reconnection during a solar flare, which is self-consistently formed and is comparable with observations in nature \citep{Wang2023a,Ren2025}.
Our systematic analysis uncovers a fundamental yet previously unresolved pattern of 3D turbulent reconnection in the flare CSs.

\begin{figure*}[htb]
\centering
\includegraphics[width=0.8\textwidth]{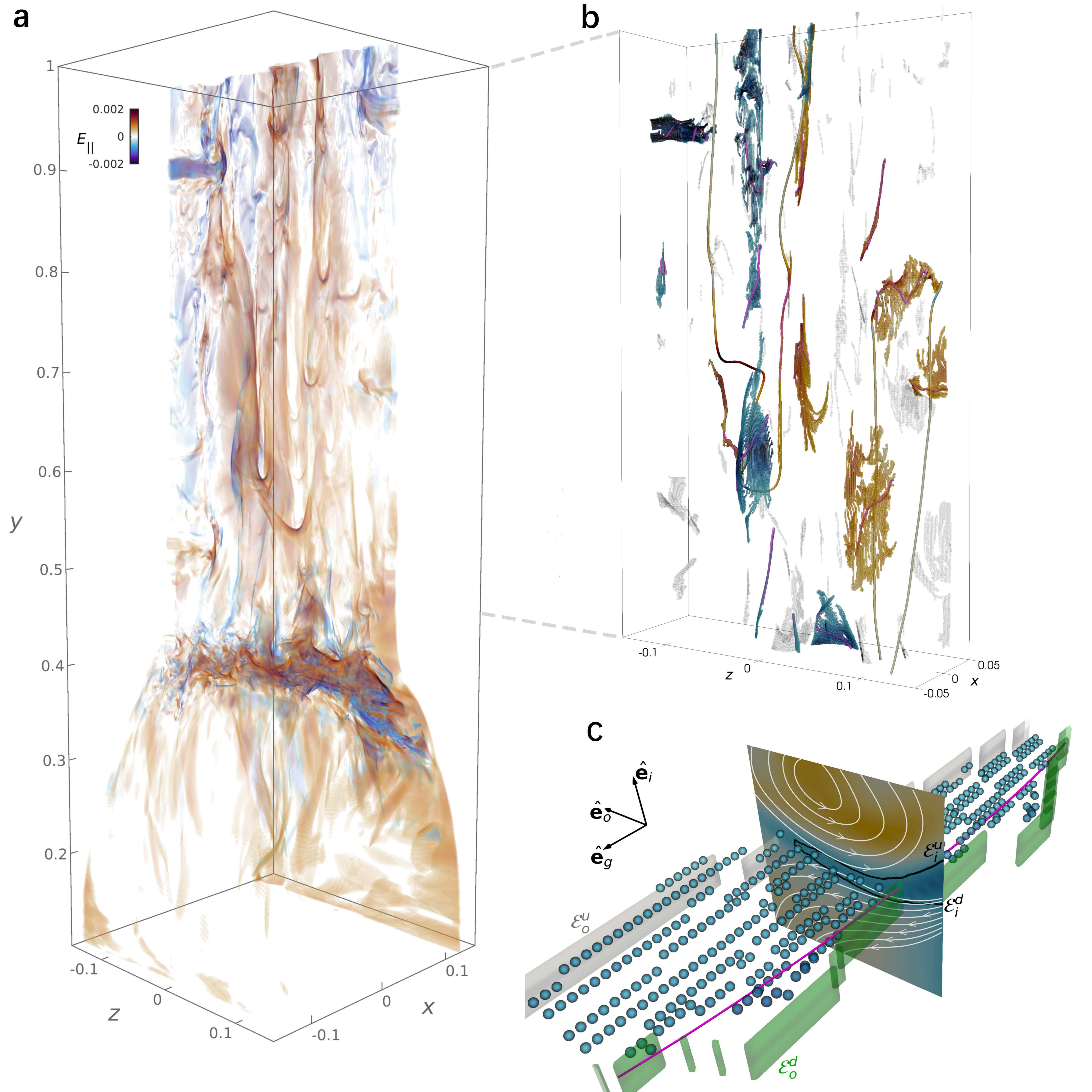}
\caption{
Diagram of reconnection regions within the turbulent flare CS.
\textbf{a} Distribution of $E_\parallel$ at $t=8.2$. Only grids with $\lvert E_\parallel\rvert \geq 5\times 10^{-4}$ are rendered for clarity.
\textbf{b} Some typical reconnection kernels.
Several reconnection kernels are highlighted and colored with $E_\parallel$.
The magenta curves denote the $\Xi_\mathrm{max}$ lines.
Two field lines connected to two $\Xi_\mathrm{max}$ lines are shown and colored by $E_\parallel$.
\textbf{c} A reconnection kernel in its intrinsic reconnection reference frame.
The dots depict the $E_\mathrm{max}$ grid points colored by $E_\parallel$.
The slice perpendicular to $\hat{\bf e}_g$ shows the 2D profile of $E_\parallel$ near the $E_\mathrm{max}$ grids, where the gray curves depict several projected field lines.
The black curves on the slice plots the contour lines of $\lvert E_\parallel\rvert=E_\mathrm{thres}$, indicating the cross sections of inflow edges $\mathcal{E}^u_i$ and $\mathcal{E}^d_i$.
The gray and green shades denote the outflow edges $\mathcal{E}^u_o$ and $\mathcal{E}^d_o$, respectively.
\label{fig:MRP_Diagram}}
\end{figure*}

\section{Simulation}
The simulation solves the resistive magnetohydrodynamics (MHD) equations incorporating necessary coronal effects (see Appendix \ref{apdx:model}).
The flare CS is $10^{8}\,\mathrm{m}$ in length and initially conforms to the standard flare model \citep{Shibata1995,Lin2015}, rooted in a high-density, low-temperature chromospheric layer \citep{Wang2021b,Wang2022,Ye2020,Shen2022}.
Fast reconnection is initiated by a localized anomalous resistivity in the corona and is subsequently dominated by a low uniform resistivity $\eta_b$, corresponding to a background Lundquist number of $S=2\times 10^5$.
A uniform mesh with a grid size of $\Delta L=26\,\mathrm{km}$ is used in the CS and flare loop top regions to guarantee the accuracy of physical results.
Under the influences of tearing mode instability (TMI), kink instability (KI), and Kelvin-Helmholtz instability (KHI) in sequence, the CS finally evolves into a self-sustained strongly turbulent state, indicated by turbulent magnetic energy comparable to background magnetic energy and a spectrum presenting an inertial region with a power-law index close to $-5/3$ \citep{Wang2023a}.
Notably, this simulation reproduces various key observational features of solar flares \citep{Wang2023a}, such as finger-like structures above the flare loop top \citep{Hanneman2014a} and turbulent loop-top regions \citep{McKenzie2013}.
Hereafter, we focus on the data within the CS region $y\in\left[0.45,1\right]$ at the final turbulent moment $t=8.2$.
In simulation, the units of length, time, velocity, electric field, magnetic field, and current density are $L_0=5\times 10^7\,\mathrm{m}$, $t_0=114.61\,\mathrm{s}$, $u_0=4.36 \times 10^5 \,\mathrm{m\,s^{-1}}$, $E_0=8.73\times 10^2\,\mathrm{V\,m^{-1}}$, $B_0 = 0.002\,\mathrm{T}$, and $J_0=3.18\times 10^{-5}\,\mathrm{A\,m^{-2}}$, respectively.

\section{Reconnection kernels}
According to the general magnetic reconnection theory, the condition for 3D reconnection is given by $\Xi\equiv \int E_\parallel\mathrm{d}s\neq 0$ \citep{Schindler1988,Hesse1988,Hesse2005,Pontin2022}, where $\Xi$ is the quasi-potential, $E_\parallel=\eta_b \bf{J\cdot B}/\lvert\bf{B}\rvert$ denotes the parallel electric field, $\bf{J}$ is the current density, $\bf{B}$ is the magnetic field, and the integration is carried out along a magnetic field line.
In a strongly turbulent state, reconnection regions in the CS characterized by intense $E_\parallel$ exhibit a highly chaotic pattern (see Figs.\ref{fig:MRP_Diagram}a and \ref{sfig_turbBlines}).
Although 3D reconnection generally occurs within a finite volume \citep{Priest2003}, among all the field lines threading this volume, there exists a special one associated with an extremal quasi-potential $\Xi_\mathrm{max}$, which corresponds to the reconnection rate \citep{Hesse2005,Wyper2015}.
For discrete numerical data, the spatial distribution of $\Xi_\mathrm{max}$ field lines can be approximately inferred from $E_\mathrm{max}$ grid points (see Appendix \ref{apdx:EmaxGrid}), identified by the criteria $\nabla_\perp E_\parallel \sim 0$ and $\lvert E_\parallel\rvert> E_\mathrm{thres}$.
Here, $\nabla_\perp$ denotes the gradient operator perpendicular to the local magnetic field, and $E_\mathrm{thres}=5\times 10^{-4}$ serves as a threshold distinguishing regions undergoing reconnection or not.
It should be noted that the statistical results presented later are robust to reasonable variations in $E_\mathrm{thres}$, as detailed in Appendix \ref{apdx:diffE}.

The $E_\mathrm{max}$ grids represent the smallest numerically resolvable units of reconnection.
To investigate the integral properties of reconnecting regions, we employ the region growing algorithm to cluster spatially connected $E_\mathrm{max}$ grids, thereby constituting the kernels of reconnection regions (Fig.\,\ref{fig:MRP_Diagram}b).
Here, ``spatially connected'' means that, for any given grid, its nearest neighbor resides within the same mesh cell.  
This approach allows the original complex turbulent reconnection structure shown in Fig.\,\ref{fig:MRP_Diagram}a to be decomposed into smaller, more analyzable components.
The reconnection kernels exhibit diverse shapes, scales, and $E_\parallel$ values (Fig.\,\ref{fig:MRP_Diagram}b).
Some extend considerable distances in specific directions, resembling the extended X-lines observed in kinetic simulations \citep{Li2023b}.
It is worth noting that the lower threshold, $E_\mathrm{thres}$, ensures all grids within a reconnection kernel have identical signs of $E_\parallel$.
Among the field lines intersecting the $E_\mathrm{max}$ grids, there is one line that can be identified with the maximum value of $\lvert\Xi\rvert$, allowing precise determination of the $\Xi_\mathrm{max}$ line for the reconnection region (see the magenta curves in Fig.\,\ref{fig:MRP_Diagram}b).
Unlike laminar reconnection, the $\Xi_\mathrm{max}$ line of a reconnection region may pass multiple reconnection regions (Fig.\,\ref{fig:MRP_Diagram}b).

For a given reconnection kernel, there exists an intrinsic reconnection frame (Fig.\,\ref{fig:MRP_Diagram}c), consisting of the guide field direction ($\hat{\bf e}_g$), the inflow direction ($\hat{\bf e}_i$), and the outflow direction ($\hat{\bf e}_o$).
$\hat{\bf e}_g$ approximately aligns with the average magnetic field of $E_\mathrm{max}$ grids.
$\hat{\bf e}_i$ is determined by the local magnetic field structures at all $E_\mathrm{max}$ grids and is approximately normal to the surface of the reconnection kernel (see Appendix \ref{apdx:frame}).
$\hat{\bf e}_o$ is set to be perpendicular to both $\hat{\bf e}_g$ and $\hat{\bf e}_i$.
Using this frame, the inflow edges ($\mathcal{E}^u_i$, $\mathcal{E}^d_i$) and outflow edges ($\mathcal{E}^u_o$, $\mathcal{E}^d_o$) surrounding a reconnection kernel can be identified (see Fig.\,\ref{fig:MRP_Diagram}c).
Here, the superscripts ``u'' and ``d'' denote the upper and down sides relative to $\hat{\bf e}_i$ or $\hat{\bf e}_o$, respectively.
The inflow edges correspond to the nearest isosurfaces with $\lvert E_\parallel\rvert=E_\mathrm{thres}$ enclosing the reconnection kernel.
The outflow edges outline the boundaries of the reconnection region at both ends of the outflow direction, with their widths along the inflow direction determined by the average thickness between inflow edges.
These edges enable us to quantitatively investigate reconnection flows and dimensionless reconnection rates of all fragmented reconnection regions.
Technical details regarding the determination of the frame and edges are elaborated in Appendix \ref{apdx:frame}.

\begin{figure*}[htb]
\centering
\includegraphics[width=0.5\textwidth]{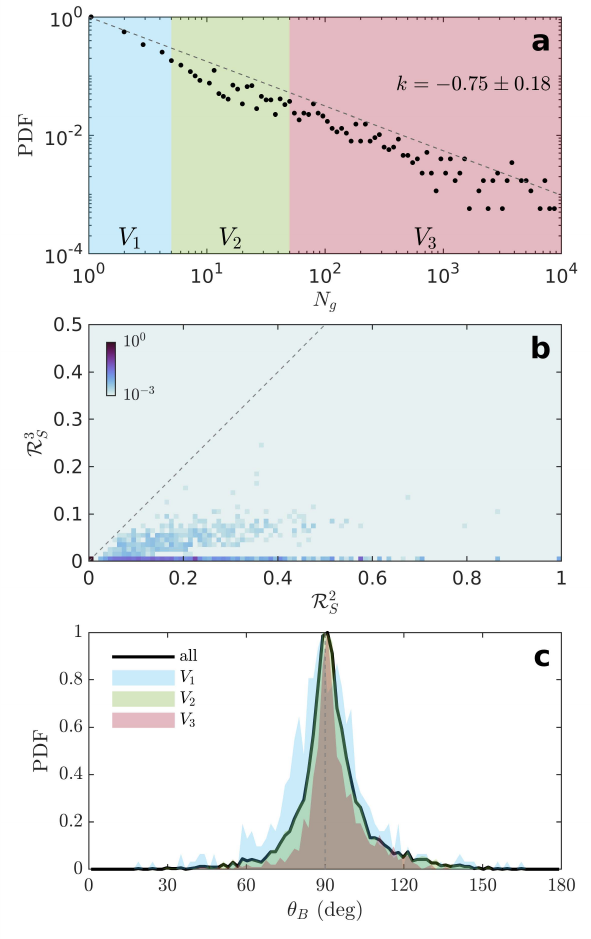}
\caption{Statistics of morphological properties of reconnection kernels.
(a) PDF of the grid numbers of reconnection kernels.
The dashed line represents the linear fit, with its slope $k$ labeled.
The colored shades $V_1$, $V_2$, and $V_3$ indicate the domains of $N_g\in[1,5)$, $[5,50)$, and $[50,\infty)$, respectively.
(b) Joint PDF of $\mathcal{R}_S^2$ and $\mathcal{R}_S^3$ for reconnection kernels with finite $\mathcal{L}_1$.
The dashed line denotes the $\mathcal{R}^3_S=\mathcal{R}^2_S$ line.
(c) PDFs of $\theta_B$ for all reconnection kernels with finite $\mathcal{L}_2$ (the black curve).
The colored shades correspond to the PDFs of $\theta_B$ for reconnection kernels of $V_1$--$V_3$ as indicated in panel (a).
All PDFs are normalized by their maximum values for clarity of comparison.
}
\label{fig:MRP_Shape}
\end{figure*}

\section{Morphological properties}
The volumes of reconnection kernels can be represented by their $E_\mathrm{max}$ grid numbers, $N_g$, which follow a power-law probability distribution function (PDF) decreasing towards larger $N_g$ (Fig.\,\ref{fig:MRP_Shape}a).
To quantitatively evaluate the shape of a reconnection kernel, we perform principal component analysis (PCA) on its grid coordinates, which outputs the coordinate variances on three principal directions denoted by $\mathcal{L}_1$, $\mathcal{L}_2$, and $\mathcal{L}_3$ in descending order.
For reconnection kernels with finite $\mathcal{L}_1$, their shapes can be evaluated by $\mathcal{R}_S^\alpha=\sqrt{\mathcal{L}_\alpha/\mathcal{L}_1}$, $\alpha=2,\,3$.
Figure \ref{fig:MRP_Shape}b shows that almost all reconnection kernels satisfy $\mathcal{R}^3_S < \mathcal{R}^2_S$, with a large portion having $\mathcal{R}^3_S$ close to 0, implying that they tend to have 2D patch-like spatial distributions.
For a reconnection kernel with $\mathcal{L}_2\neq 0$, the characteristic vector $\hat{\bf e}_c$ corresponding to the smallest variance $\mathcal{L}_3$ is approximately perpendicular to the surface of the reconnection kernel.
The mean angle between $\hat{\bf e}_c$ and the magnetic field at $E_\mathrm{max}$ grids $\mathbf{B}_{gl}$, evaluated by $\theta_B=\langle\arccos(\hat{\bf b}_{gl}\cdot \hat{\bf e}_c)\rangle$, concentrates around $90^\circ$ (Fig.\,\ref{fig:MRP_Shape}c), indicating that reconnection kernels tend to form along the local magnetic flux surfaces.
Here $\hat{\bf b}_{gl}$ is the unit vector of $\mathbf{B}_{gl}$ and $\langle \cdot \rangle$ in this paper denotes averaging over all $E_\mathrm{max}$ grids of a reconnection kernel.

\begin{figure*}[htb]
\centering
\includegraphics[width=0.8\textwidth]{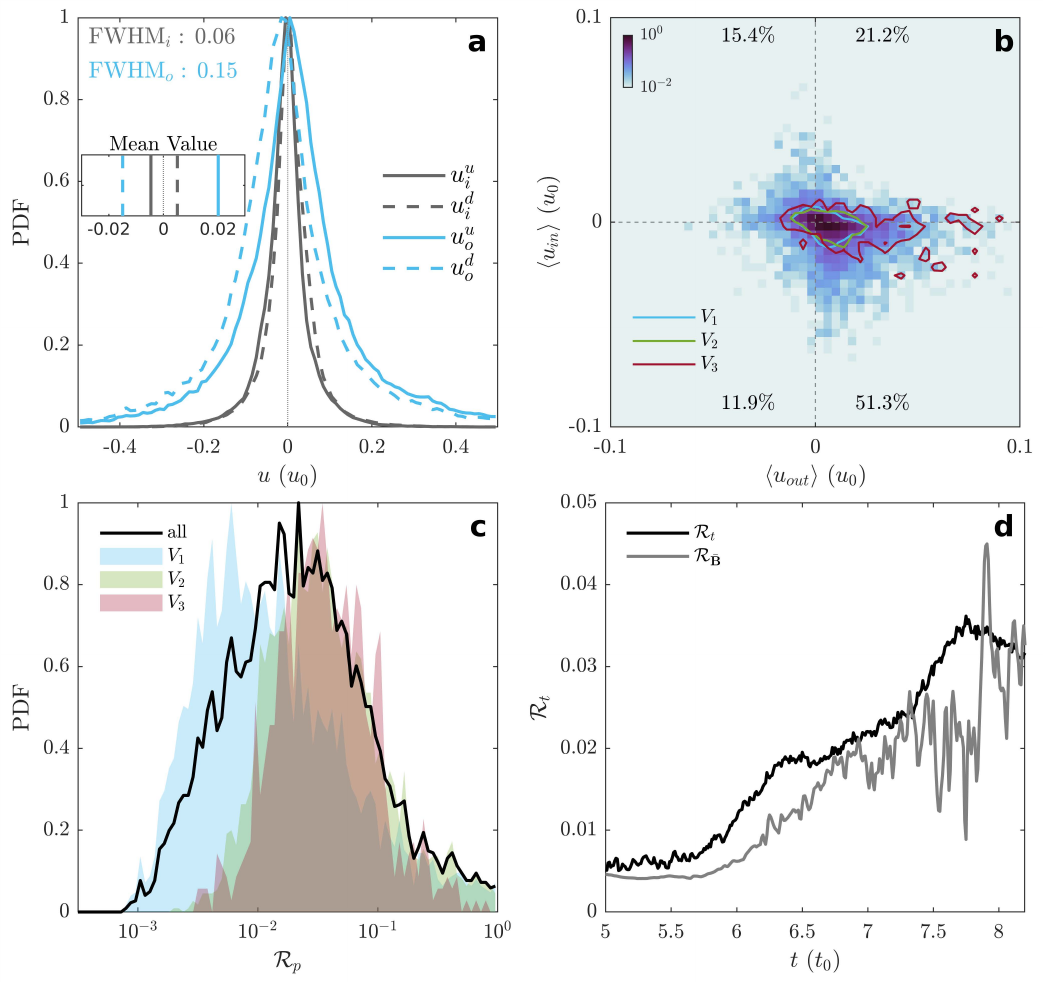}
\caption{Statistics of reconnection properties.
(a) PDFs of $u^u_i$, $u^d_i$, $u^u_o$, and $u^d_o$.
$\mathrm{FWHM}_i$ represents the averaged FWHM of the $u^u_i$ and $u^d_i$ distributions, while $\mathrm{FWHM}_o$ is the FWHM for outflow velocities.
The mean values of the four PDFs are denoted by the zoom-in window.
(b) Joint PDF of $\langle u_{out} \rangle$ and $\langle u_{in} \rangle$ for all reconnection kernels.
The sampling percentages in each quadrant are labeled.
Three color curves denote the contours of $\mathrm{PDF} = 0.2$ for reconnection kernels in $V_1$, $V_2$, and $V_3$, respectively.
(c) PDF of the reconnection rate $\mathcal{R}_p$.
The black curve shows the result for all reconnection kernels, while the colored shades correspond to the different $V_1$--$V_3$ categories as indicated in Fig.\,\ref{fig:MRP_Shape}a.
(d) Temporal evolution of the total reconnection rates evaluated by $\mathcal{R}_t$ and $\mathcal{R}_{\bar{\bf B}}$.
}
\label{fig:MRP_V_Xi}
\end{figure*}

\section{Reconnection properties}
The reconnection inflows and outflows associated with a reconnection kernel can be determined at its inflow and outflow edges (Fig.\,\ref{fig:MRP_Diagram}c).
Specifically, the inflow velocity on the upper inflow edge $\mathcal{E}^u_i$ is calculated as $u^{u}_i = \hat{\bf e}_i \cdot [\mathbf{u}(\mathcal{E}_i^{u}) - \langle \mathbf{u} \rangle]$, and $u^d_i$, $u^u_o$, and $u^d_o$ at the other three edges can be obtained similarly.
Generally speaking, the average velocities of reconnection kernels ($\langle\mathbf{u}\rangle$) have finite values, representing their global motions.
We collect the four velocities sampled at the edge grids for all reconnection kernels to obtain their PDFs, which exhibit the flow velocity distributions at the inflow and outflow edges (Fig.\,\ref{fig:MRP_V_Xi}a). 
It is found that all four PDFs exhibit significant broadening and approximate symmetry with respect to their peaks (Fig.\,\ref{fig:MRP_V_Xi}a).
Their full widths at half maximums (FWHMs) are much larger than their mean values, indicating a strong turbulent regime. 
The broadening of PDFs reflects the inherent uncertainties in determining reconnection flows of the fragmented reconnection regions, presenting a distinct nature compared with traditional laminar reconnection with deterministic reconnection flows.

On average, the PDF of the upper inflow velocities drifts towards negative values, while the down inflow velocities tend towards positive ones, reflecting an inward flow along $\hat{\bf e}_i$; conversely, the drift directions of the velocities at the upper and down outflow edges are opposite to those of inflows, indicating an outward flow along $\hat{\bf e}_o$ (Fig.\,\ref{fig:MRP_V_Xi}a).
The mean inflow and outflow velocities for each reconnection kernel are given by $\langle u_{in} \rangle = \langle u^{u,j}_i \rangle - \langle u^{d,j}_i \rangle$ and $\langle u_{out} \rangle = \langle u^{u,j}_o \rangle - \langle u^{d,j}_o \rangle$, respectively.
Their joint PDF shows that reconnection kernels are most probably located in the fourth quadrant with $\langle u_{in} \rangle < 0$ and $\langle u_{out} \rangle > 0$ (Fig.\,\ref{fig:MRP_V_Xi}b).
$\langle u_{in} \rangle < 0$ means that the plasmas at inflow edges move towards the reconnection kernel, while $\langle u_{out} \rangle > 0$ indicates that the plasmas are expelled out across the outflow edges.
This average effect is consistent with the flow patterns surrounding the reconnection region in Sweet-Parker and Petschek models \citep{Priest2000}.
However, it is still possible for a reconnection kernel to enter the other three quadrants with a counter-intuitive flow pattern (Fig.\,\ref{fig:MRP_V_Xi}b), which could be a new feature for 3D turbulent reconnection.

The reconnection rate of each reconnection kernel is given by $\Xi_\mathrm{max}$ \citep{Hesse2005}.
Its dimensionless value can be obtained by $\mathcal{R}_p = \lvert\Xi_\mathrm{max}\rvert / (B_{in} V_{Ain} L_{in})$.
Here, $B_{in}$ and $V_{Ain}$ are the inflow parameters calculated by the average magnetic field and Alfv\'{e}nic velocity at the two inflow edges $\mathcal{E}^u_i$ and $\mathcal{E}^d_i$, and $L_{in}$, reflecting the length of CS analogous to the 2D model, is set as the standard deviation of the reconnection kernel along $\hat{\bf e}_o$.
Similar to the velocities, the PDF of $\mathcal{R}_p$ also exhibits a strong broadening with the main body in the range of $0.01$--$0.1$ (see the black curve in Fig.\,\ref{fig:MRP_V_Xi}c).

At a given moment $t$, the total reconnection rate contributed by all fragmented reconnected regions is given by $\mathcal{R}_t = \sum \lvert\Xi_\mathrm{max}\rvert / (B_0 u_{0} L_0)$ (Fig.\,\ref{fig:MRP_V_Xi}d), where the summation is taken over all reconnection kernels.
Traditionally, for a CS with approximate translation symmetry along the guide field direction, the reconnection rate $\mathcal{R}_{\bar{\bf B}}$ can be evaluated from the 2D magnetic field averaged along the guide field direction, $\bar{\bf B} = \langle \mathbf{B} \rangle_z$ \citep{Wang2023a, Huang2016}.
The mean-field approach relies on identifying the principal X-point of $\bar{\bf B}$, which typically exhibits spatial jumps under turbulent conditions, leading to significant temporal fluctuations (see the gray curve in Fig.\,\ref{fig:MRP_V_Xi}d). 
However, by including the contributions from all reconnection fragments, the evolution of $\mathcal{R}_t$ shows considerably smaller fluctuations and follows a global trend similar to $\mathcal{R}_{\bar{\bf B}}$ (see the dark curve in Fig.\,\ref{fig:MRP_V_Xi}d).
The curve of $\mathcal{R}_t$ exhibits two rising stages corresponding to the development of TMI during $6 < t < 6.5$ and the growth of turbulence after $t = 7$, respectively.
Furthermore, $\mathcal{R}_t$ is found to be systematically larger than $\mathcal{R}_{\bar{\bf B}}$, which shows that $\mathcal{R}_{\bar{\bf B}}$, as adopted previously, tends to underestimate the total reconnection rate.

\begin{figure*}[htb]
\centering
\includegraphics[width=0.6\textwidth]{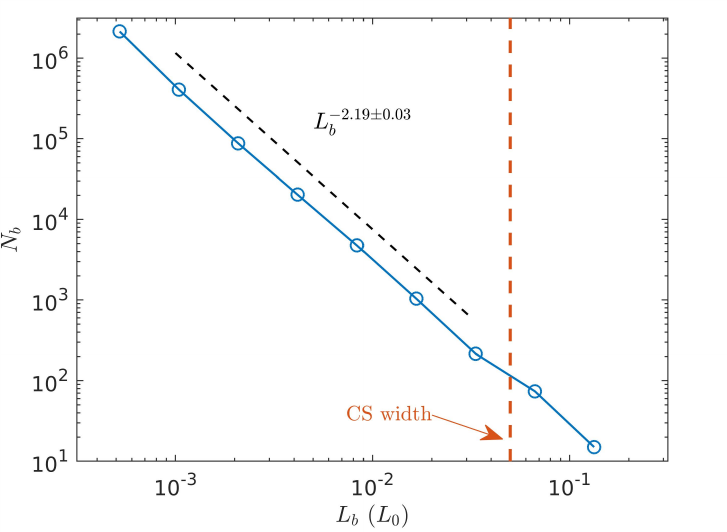}
\caption{Fractal analysis of 3D regions covering all reconnection kernels.
The blue circles depict the result samples from box-counting analysis.
The orange dashed line indicates the maximum $z$-direction width of the CS region at $t=8.2$ (Fig.\,\ref{fig:MRP_Diagram}a and also see \cite{Ren2025}).
The black dashed line represents the power-law index obtained by linear fit of samples with box sizes smaller than the CS width.
}
\label{fig:DF}
\end{figure*}

\section{Scale-independency}
To investigate the influences of reconnection kernel scales on statistical results, we categorize the reconnection kernels into three groups: $V_1$ ($N_g<5$), $V_2$ ($5\leq N_g<50$), and $V_3$ ($N_g\geq 50$) (Fig.\,\ref{fig:MRP_Shape}a).
The PDF of $\theta_B$ remains nearly unchanged across different reconnection kernel sizes (Fig.\,\ref{fig:MRP_Shape}c).
For the joint PDFs of $\langle u_{in} \rangle$ and $\langle u_{out} \rangle$, the proportions in the fourth quadrant ($\langle u_{in} \rangle < 0$ and $\langle u_{out} \rangle > 0$) are similar across all categories, with values of $50.4\%$, $52.2\%$, and $52.5\%$ for $V_1$, $V_2$, and $V_3$, respectively (Fig.\,\ref{fig:MRP_V_Xi}b).
The PDFs of $\mathcal{R}_p$ for the three categories exhibit similar trends (Fig.\,\ref{fig:MRP_V_Xi}c).
For $V_2$ and $V_3$, the PDFs almost overlap; for $V_1$, the peak shifts towards smaller values, but 41.7\% of the samples still fall within the range of $\left[0.01,0.1\right]$. 
It should be noted that the reconnection kernels in $V_1$ contain fewer than 5 grids, approaching the resolution limit of the simulation, and are likely influenced by numerical errors.
Nevertheless, the statistical trends observed in $V_1$ remain highly similar to those of the larger reconnection kernels.
Therefore, the morphologies, flows, and rates of reconnection kernels can be considered approximately scale-independent.

We also perform a standard 3D box-counting analysis to determine the fractal behavior of the regions covering all reconnection kernels (see Appendix \ref{apdx:fd} for technical details).
The relationship between the number of covering boxes ($N_b$) and their size ($L_b$) exhibits a power-law distribution, yielding a fractal dimension of $D_F = 2.19$.
This result suggests that the reconnection kernels exhibit a fractal structure in 3D space with a relatively low filling factor, further supporting the scale-independent quasi-2D patten of reconnection kernels (Fig.\,\ref{fig:DF}).

\section{Summary and Discussion}\label{discussion}
A novel approach is developed to quantitatively analyze 3D reconnection within strongly turbulent flare CS.
We recognize multitudes of fragmented reconnection kernels and disclose their basic properties including the morphologies, reconnection in/out-flows, and reconnection rates, as well as the corresponding statistical distributions.

It is proved that the reconnection kernels tend to have a patch-like structure aligned with local magnetic flux surface.
For the first time, it is shown that, due to strong turbulence, the fragmented reconnection regions produce fluctuated in/out-flows and reconnection rates, represented by the significant broadenings in their PDFs.
Despite a strong fluctuation, on average, they most probably induce flows compressing plasmas towards the reconnection kernels and expel exhausts in the outflow direction.
The reconnection rates mainly take values of 0.01--0.1, coinciding with the values derived by previous studies for different reconnection configurations \citep{Cassak2017}.
More importantly, the statistical laws are found to be approximately independent of the reconnection kernel scales, indicating the existence of a fundamental reconnection pattern.
At last, we propose a new strategy to evaluate the reconnection rate of the highly fragmented 3D turbulent reconnection.
Compared with the mean-field method which only works for 3D systems with approximate translation symmetry on the guide field direction \citep{Huang2016}, the new method can be applied to arbitrary 3D systems and estimate the reconnection rate $\mathcal{R}_t$ more accurately, thus assessing its intrinsic evolution.

\begin{figure*}[htb]
\centering
\includegraphics[width=0.7\textwidth]{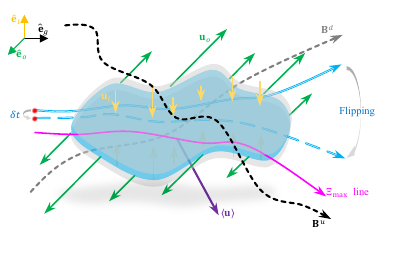}
\caption{ 
Schematic of the 3D reconnection pattern of an arbitrary reconnection fragment within strongly turbulent CS.
The blue surface denotes the reconnection kernel, sandwiched between two gray surfaces indicating the inflow edges.
The reconnection kernel has a global velocity as marked by $\langle\bf u\rangle$.
The yellow and green arrows denote the inflow and outflow velocities, respectively, and their lengths indicate the speed strengths.
The dark and light dashed curves plot the background field lines migrated toward the reconnection kernel by inflows.
The magenta arrow curve denotes the $\Xi_\mathrm{max}$ line.
The blue solid curve plots a field line passing through the dissipation region, which is traced from a small plasma element denoted by a red dot.
After a short time $\delta t$, the red dot flows a small distance, while its field line, illustrated by a blue dashed curve, flips significantly on the other end as affected by the reconnection within the reconnection region.}
\label{fig:Cartoon}
\end{figure*}

A scenario of 3D reconnection within strongly turbulent flare CSs is suggested and illustrated by Fig.\,\ref{fig:Cartoon}.
The reconnection kernel with an irregular shape tends to have a 2D patch-like configuration.
Owning to the turbulence, the reconnection inflows and outflows are violently perturbed (see the velocity arrows in Fig.\,\ref{fig:Cartoon}).
The $\Xi_\mathrm{max}$ line, a special field line threading the reconnection kernel, satisfies the line conservation \citep{Hesse2005,Wyper2015} and approximately moves along with the reconnection kernel (see the magenta curve in Fig.\,\ref{fig:Cartoon}).
If a field line passes through the diffusion region enclosing the reconnection kernel, its start point flows with the plasma (see the red dot in Fig.\,\ref{fig:Cartoon}) and moves a small distance after a short time $\delta t$ but its endpoint might flip a large distance (see the blue curves in Fig.\,\ref{fig:Cartoon} or Fig.\,7 of \cite{Priest2003}).
This flipping is similar to the slipping of flare loops as frequently observed along flare ribbons \citep{Aulanier2006,Aulanier2007}.
Thousands of reconnection fragments with various flow patterns and reconnection rates constitute the complex reconnection process in a turbulent CS, but their statistical distributions are independent of scales.
The reconnection kernels collectively exhibit a fractal structure in 3D space with a fractal dimension greater than 2.
This effectively generalizes the concept of fractal reconnection by \cite{Shibata2001} to 3D turbulent reconnection and provides a quantitative validation.

The quasi-separatrix layers (QSLs) are commonly used to identify where the reconnection may occur \citep{Pontin2022,Pontin2024a}, but are not able to pinpoint the reconnection \citep{Reid2020}.
In contrast, the methodology developed here not only precisely locates the reconnection sites but also quantitatively determines their physical properties.
Furthermore, while our analysis targets the collisional reconnection regime, the method that is developed from general reconnection theory and only requires magnetic field, non-ideal electric field, and velocities as inputs can be applied to kinetic simulations of collisionless reconnection.
The resulting statistics of local reconnection properties can aid in initializing physical parameters for small-scale kinetic or hybrid simulations and provide strategies for particle injection in MHD-particle models \citep{Zhang2024,Drake2019,Arnold2021,Sun2023}.

The CS for eruptive flares in our simulation has a special configuration, with magnetic field lines rooted in the solar photosphere and extending outward.
The guide field within the CS is aligned along the polarity inversion line  \citep{Janvier2014,Xing2024} and varies spatially and temporally in magnitude \citep{Aulanier2012,Dahlin2022}.
This distinctive configuration plays a key role in determining the development of the turbulent CS and in forming fine flare features observed.
Furthermore, despite the turbulent nature of the reconnecting CS varying with time in our simulation, the statistical properties derived above exhibit a limited variation (Fig.\,\ref{sfig_diffTime_shapeXi}), suggesting the existence of a unified pattern for 3D turbulent reconnection in flare CSs. Nevertheless, more investigations are still encouraged to justify such a fundamental pattern for other reconnection regimes; the methodology developed here offers a valuable analysis tool.

\appendix
\restartappendixnumbering

\section{Numerical model}\label{apdx:model} %

We utilize the simulation data from \cite{Wang2023a}, with its main configurations summarized here for completeness.
The \textsf{Athena++} code is employed to solve the resistive MHD equations \citep{Stone2020}, incorporating the effects of anisotropic thermal conduction, radiative cooling, background heating, and solar gravity.
Thermal conduction is assumed to be aligned with the magnetic field, with the conductivity coefficient determined by the classical Spitzer model \citep{Yokoyama2001}.
Radiative cooling is considered optically thin and is computed using a widely adopted model in solar simulations \citep{Klimchuk2008,Ye2020,Wang2022}.
Background heating is configured to balance the cooling effects at the initial moment \citep{Ye2020,Ni2015}.

\begin{figure}[htb]
\centering
\includegraphics[width=0.8\textwidth]{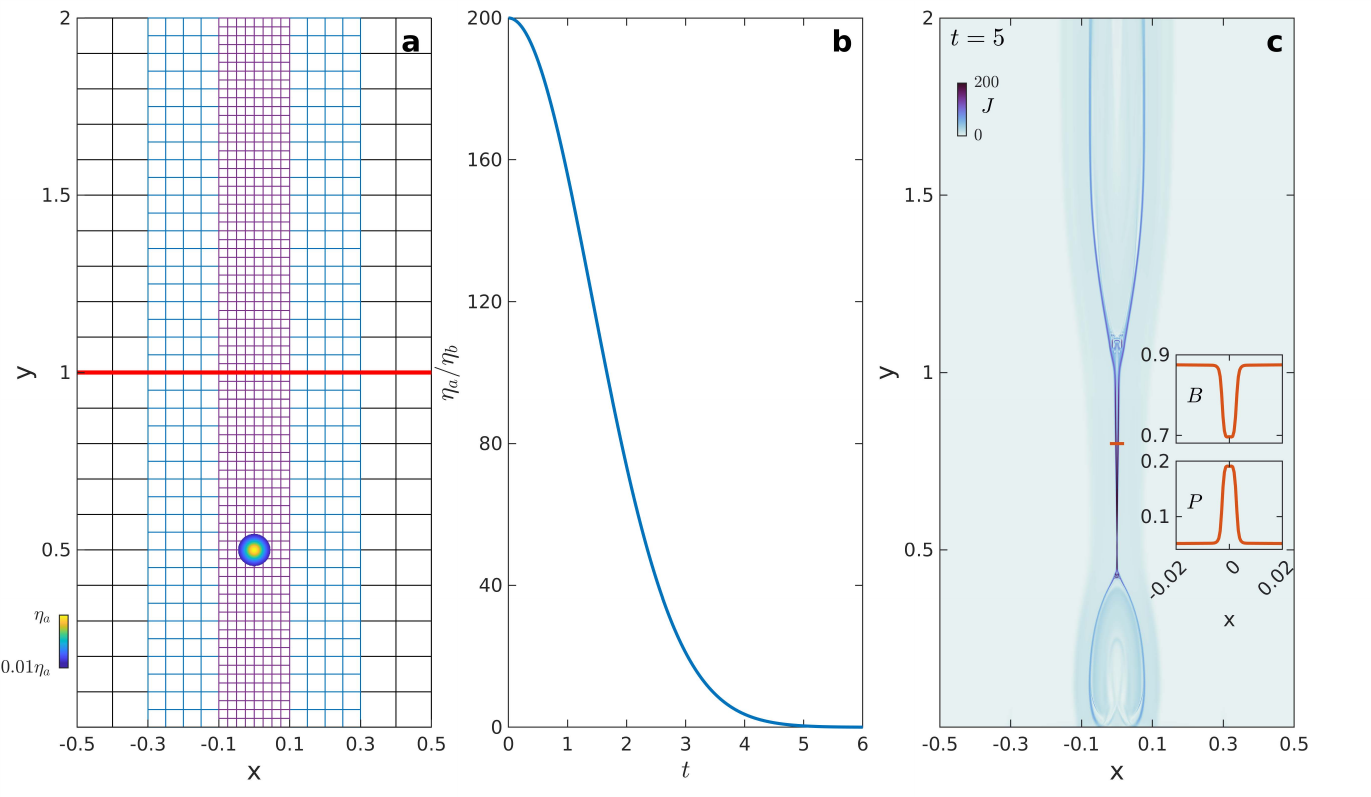}
\caption{
\textbf{a} Schematic representation of the 3-level static mesh refinement utilized in the simulation.
The initial spatial distribution of the anomalous resistivity $\eta_a$ is provided.
Grid levels 0, 1, and 2 are represented by black, purple, and blue colors, respectively, with the highest resolution grid points at level 2 achieving a spatial resolution of $26,\mathrm{km}$.
\textbf{b} Temporal evolution of the maximum value of $\eta_a$.
\textbf{c} Current density distribution at $t=5$.
The profiles of magnetic field strength ($B$) and thermal pressure ($P$) along a horizontal slit are also shown, highlighting the presence of slow shocks at the boundaries of the current sheet.
\label{sfig_meshgrid}}
\end{figure}

\begin{figure}[htb]
\centering
\includegraphics[width=0.8\textwidth]{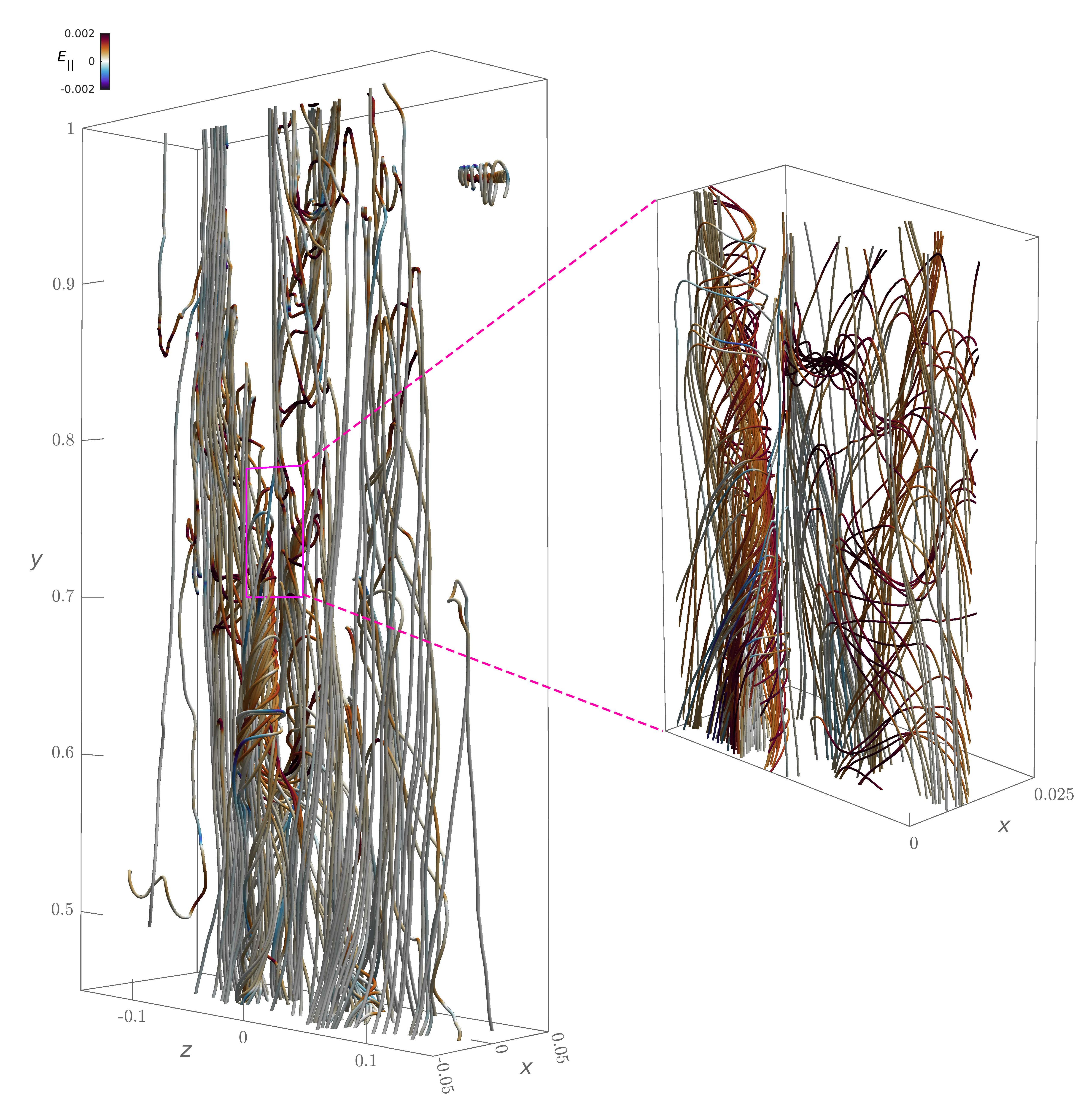}
\caption{Diagram of chaotic magnetic field lines within the turbulent flare CS at $t=8.2$.
The initial positions for field-line tracing are randomly sampled in regions with $\lvert E_\parallel\rvert>7.5\times 10^{-4}$.
\label{sfig_turbBlines}}
\end{figure}

The initial atmosphere is in hydrostatic equilibrium, maintained by the balance between thermal pressure and gravity.
It consists of a $1\,\mathrm{MK}$ corona and a low-temperature, high-density chromospheric layer at the base.
While thermal conduction can influence the thermal balance at the transition layer between the chromosphere and the corona, where a steep temperature gradient exists, its effect is negligible compared to the dominant fast reconnection process.
Consequently, the background atmosphere remains nearly static throughout the simulation timescale.
To prevent numerical instabilities at the lower boundary, the radiative cooling and background heating terms are tapered to zero below the coronal region.

The initial CS is localized in the region $y\in\left[-0.1,0.1\right]$, formed by a force-free magnetic field with a guide field in the $z$-direction and opposite $y$ components across the $x=0$ plane.
The background resistivity, $\eta_b$, is set to a low value of $5\times 10^{-6}$ to achieve a high Lundquist number condition.
To initiate fast reconnection, a localized anomalous resistivity, $\eta_a$, is introduced at $y=0.5$ (see Fig.\,\ref{sfig_meshgrid}a).
If $\eta_a$ were constant over time, the resulting evolution would lead to a standard Petschek-type reconnection, suppressing the formation of plasmoids in the CS region \citep{Shibata2023}.
In our simulation, however, $\eta_a$ is allowed to decay temporally, nearly vanishing by $t=5$, thus leaving the subsequent evolution dominated by the background resistivity (Fig.\,\ref{sfig_meshgrid}b).
This setup produces a long-stretched CS at $t=5$, closely resembling the structure self-consistently formed by solar eruptions (Fig.\,\ref{sfig_meshgrid}c and also see \cite{Dahlin2022}).
At this moment, the CS exhibits a thickness that is nearly uniform along the $y$-direction and presents typical slow shocks at its boundaries (see Fig.\,\ref{sfig_meshgrid}c).
Subsequently, the CS continues to elongate as the upper principal plasmoid propagates upward, maintaining this configuration until the onset of TMI around $t=6$.
The evolution prior to $t=5$ provides the initial condition for the later fast reconnection and is therefore excluded from our analysis.

At $z=0$, a symmetric boundary condition is imposed \citep{Yokoyama2001}, while all other boundaries are set as open boundaries using equal-value extrapolation.
To prevent numerical inflow at the upper boundary, the $y$-direction velocity is set to zero if it becomes negative.
To minimize the influence of the upper free boundary, only data below $y=1$ are used in the analysis.
Since the main reconnection process and associated turbulence occur within the CS region, a uniform mesh with a grid size of $26\,\mathrm{km}$ is applied in this region to ensure accuracy.
Outside the CS region, the grid size is expanded by two levels using the static mesh refinement technique to reduce computational costs (see Fig.\,\ref{sfig_meshgrid}a).
The simulation domain is defined as $-0.5\leq x\leq 0.5$, $0\leq y\leq 2$, and $-0.15\leq z\leq 0.15$, corresponding to an effective mesh resolution of $1920 \times 3840 \times 576$.
The accuracy of main results has been verified by convergence tests \citep{Wang2023a}.

For the conservation part of MHD equations, the HLLD Riemann solver is employed to minimize numerical resistivity \citep{Miyoshi2005}.
A second-order piecewise linear method is utilized for spatial reconstruction, while the time evolution is computed using the second-order van Leer predictor-corrector scheme.
Source terms are handled via the explicit operator-splitting method.
The anisotropic thermal conduction is solved using a slope-limited asymmetric approach \citep{Sharma2007}, with the second-order RKL2 super-time-stepping algorithm applied to reduce computational costs \citep{Meyer2014}.

The simulation ends at $t=8.2$, corresponding to a physical time of $15.66\,\mathrm{min}$.
At this moment, the CS enters a strongly turbulent state, and the magnetic field lines exhibit a highly chaotic pattern (see Fig.\,\ref{sfig_turbBlines}).

\section{Physical meaning of $E_\mathrm{max}$ grids}\label{apdx:EmaxGrid} %
The $\Xi_\mathrm{max}$ line threading a reconnection region satisfies $\partial\Xi/\partial\alpha = \partial\Xi/\partial\beta = 0$ \citep{Hesse2005,Wyper2015}, where $\alpha$ and $\beta$ are the Euler potentials.
Given that ${\bf B} = \nabla\alpha \times \nabla\beta$, the operators $\partial/\partial\alpha$ and $\partial/\partial\beta$, which act across different field lines, are equivalent to $\nabla_\perp$ under the local approximation of a very short field line.
For regions with finite $E_\parallel$, a sufficient condition for a field line to exhibit an extremal $\Xi$ is that all locations along the field line where $E_\parallel \neq 0$ satisfy $\nabla_\perp E_\parallel = 0$, which can be verified as follows:
\begin{equation}
   0=\int \nabla_\perp E_\parallel\mathrm{d}s=\nabla_\perp\int E_\parallel\mathrm{d}s=\nabla_\perp\Xi\,.
\end{equation}
Hence, although it is not a necessary condition, the distribution of $E_\mathrm{max}$ can serve as an approximation for $\Xi$ lines.
For discrete numerical data, the $E_\mathrm{max}$ grids can be identified using the procedure outlined by \cite{Wang2024}.
By selecting these $E_\mathrm{max}$ grids, the complexity of the original system is significantly reduced, and computational efficiency is enhanced without compromising the principal information about reconnection.

\section{Local reconnection frame and In/out-flow edges}\label{apdx:frame} %
\begin{figure}[htb]
\centering
\includegraphics[width=0.8\textwidth]{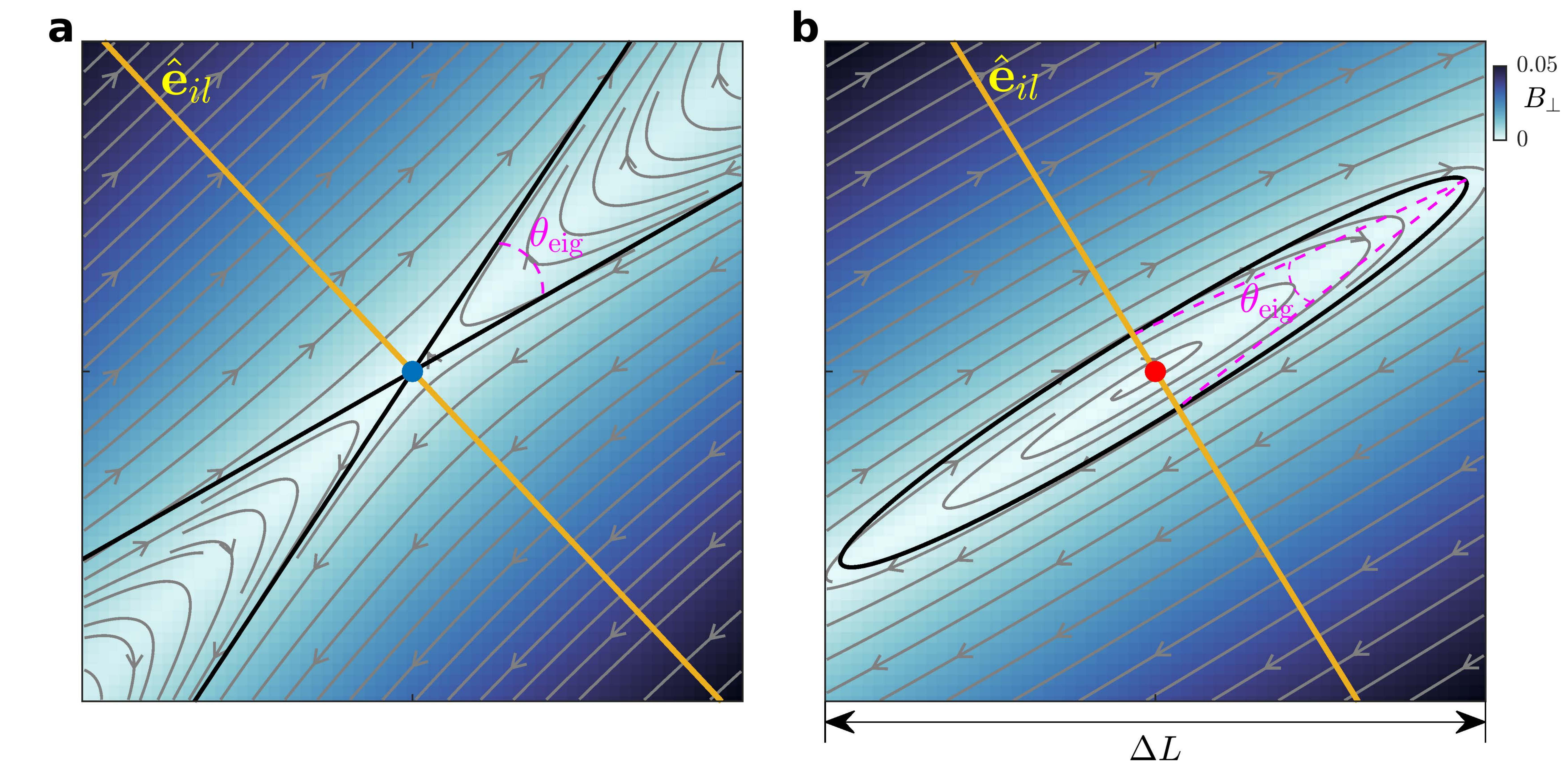}
\caption{Local magnetic structures of typical X- (a) and O-type (b) $E_\mathrm{max}$ grids.
The background color represents the strength of $\mathbf{B}_\perp$, with field lines denoted by gray arrow curves.
The thick black curves outline the separatrix lines and the elliptical field line of $\mathbf{B}'_\perp$ for $P_X$ and $P_O$, respectively.
Definitions of $\theta_\mathrm{eig}$ are denoted by magenta markers.
The yellow lines depict the directions of $\hat{\bf e}_{il}$.
\label{sfig_diagram_XO}}
\end{figure}

Unlike 2D reconnection, a globally constant guide field direction might be unavailable in 3D reconnection, especially in systems with strong turbulence \citep{Wang2024}. 
However, the magnetic field of an $E_\mathrm{max}$ grid intrinsically defines its local guide field $\mathbf{B}_{gl}$.
Additionally, the local magnetic structures near this $E_\mathrm{max}$ grid can be examined by $\mathbf{B}_\perp$, the 2D component lying on the projection plane perpendicular to $\mathbf{B}_{gl}$ (Fig.\,\ref{sfig_diagram_XO}).
Theoretically, $\mathbf{B}_\perp$ can be categorized into nine types (see Tab.\,1 in \cite{Wang2024}).
Within the CS region, the 3D X- and O-type grids are found to dominate, accounting for 41.0\% and 51.3\% in total, respectively.
The rest are 3D repelling/attracting grids.

The anisotropic properties of $\mathbf{B}_\perp$ are represented by its source-free part $\mathbf{B}'_\perp=\mathbf{B}_\perp-(\nabla_p\cdot \mathbf{B}_\perp)\mathbf{R}/2$, where $\nabla_p$ and $\mathbf{R}$ denote the 2D gradient operator and coordinate vector on the projection plane, respectively \citep{Wang2024}.
For both X- and O-type $\mathbf{B}'_\perp$, there exists an eigen-angle $\theta_\mathrm{eig}$ (see Fig.\,\ref{sfig_diagram_XO} and Eq.\,23 in \cite{Wang2024}), which reflects the local exhaust opening angle analogous to 2D reconnection \citep{Cassak2017,Liu2022d}. 
The bisectors of $\theta_\mathrm{eig}$ lie along the directions of maximum curvature of $\mathbf{B}'_\perp$.
The line perpendicular to the open direction of $\theta_\mathrm{eig}$, denoted by $\hat{\bf e}_{il}$, corresponds to the direction of strong shear of $\mathbf{B}'_\perp$, which defines the local intrinsic inflow direction (Fig.\,\ref{sfig_diagram_XO}).

\begin{figure}[htb]
\centering
\includegraphics[width=0.7\textwidth]{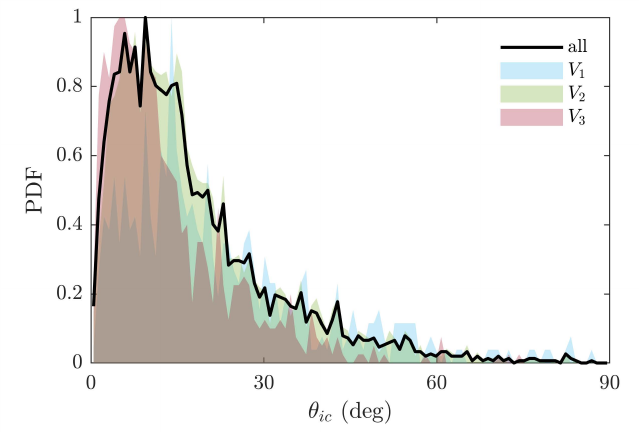}
\caption{PDFs of $\theta_{ic}$ for all reconnection kernels with finite $\mathcal{L}_2$.
The colored shades correspond to the PDFs of $V_1$--$V_3$ as indicated in Fig.\,\ref{fig:MRP_Shape}a.
\label{sfig_theta_ic}}
\end{figure}

For a reconnection kernel, the intrinsic reconnection frame can be determined by its average directions of $\hat{\bf e}_{il}$ and $\mathbf{B}_{gl}$ as follows.
First, the origin point is set as the mean position of $E_\mathrm{max}$ grids.
Second, the inflow direction is evaluated by $\hat{\bf e}_i=\langle\hat{\bf e}_{il}\rangle/\lvert\langle\hat{\bf e}_{il}\rangle\rvert$.
Before averaging, we let $\hat{\bf e}_{il}$ point to a positive direction defined by $\hat{\bf e}_c$, namely, $\hat{\bf e}_{il}\cdot \hat{\bf e}_c \ge 0$.
For reconnection kernels lacking a valid $\hat{\bf e}_c$, the positive direction can be set arbitrarily.
According to Fig.\,\ref{sfig_theta_ic}, the angle $\theta_{ic}$ spanned by $\hat{\bf e}_i$ and $\hat{\bf e}_c$ is most probably smaller than $30^\circ$, indicating that the inflow direction approximately aligns with the direction across the surfaces of reconnection kernels.
Moreover, the PDF of $\theta_{ic}$ is also independent of reconnection kernel scales.
Third, the outflow direction should be perpendicular to the directions of inflow and guide field and thus is evaluated by $\hat{\bf e}_o = \langle\hat{\bf B}_{gl}\rangle \times \hat{\bf e}_i/\lvert\langle\hat{\bf B}_{gl}\rangle \times \hat{\bf e}_i\rvert$. 
Fourth, to complete the orthogonal reference frame, the guide field direction is modified as $\hat{\bf e}_g = \hat{\bf e}_i \times \hat{\bf e}_o$.

To find the inflow edges of a reconnection kernel, at the $j$th $E_\mathrm{max}$ grid $\mathbf{r}_0^j$, we set up a slit parallel to $\hat{\bf e}_i$ and locate the nearest positions on both sides satisfying $\lvert E_\parallel\rvert < E_\mathrm{thres}$, denoted as $\mathbf{r}_i^{u,j}$ and $\mathbf{r}_i^{d,j}$.
The sets of $\mathbf{r}_i^{u,j}$ and $\mathbf{r}_i^{d,j}$ outline the upper and down inflow edges $\mathcal{E}_i^u$ and $\mathcal{E}_i^d$, respectively.
The thickness between $\mathcal{E}_i^u$ and $\mathcal{E}_i^d$ at $\mathbf{r}_0^j$ is given by $W^j_p = (\mathbf{r}_i^{u,j} - \mathbf{r}_i^{d,j}) \cdot \hat{\bf e}_i$.
To locate the outflow edges, $\mathcal{E}_o^u$ and $\mathcal{E}_o^d$, we first outline the boundary lines of a reconnection kernel at both ends along the $\hat{\bf e}_o$ direction on the $\hat{\bf e}_o$-$\hat{\bf e}_g$ plane.
Then, the boundary lines are symmetrically extended along $\pm \hat{\bf e}_i$ to form the two outflow edges. 
The extension lengths along $\hat{\bf e}_i$ are determined by the mean thickness between the inflow edges, given by $W_p = \langle W^j_p \rangle$.

\begin{figure}[htb]
\centering
\includegraphics[width=1\textwidth]{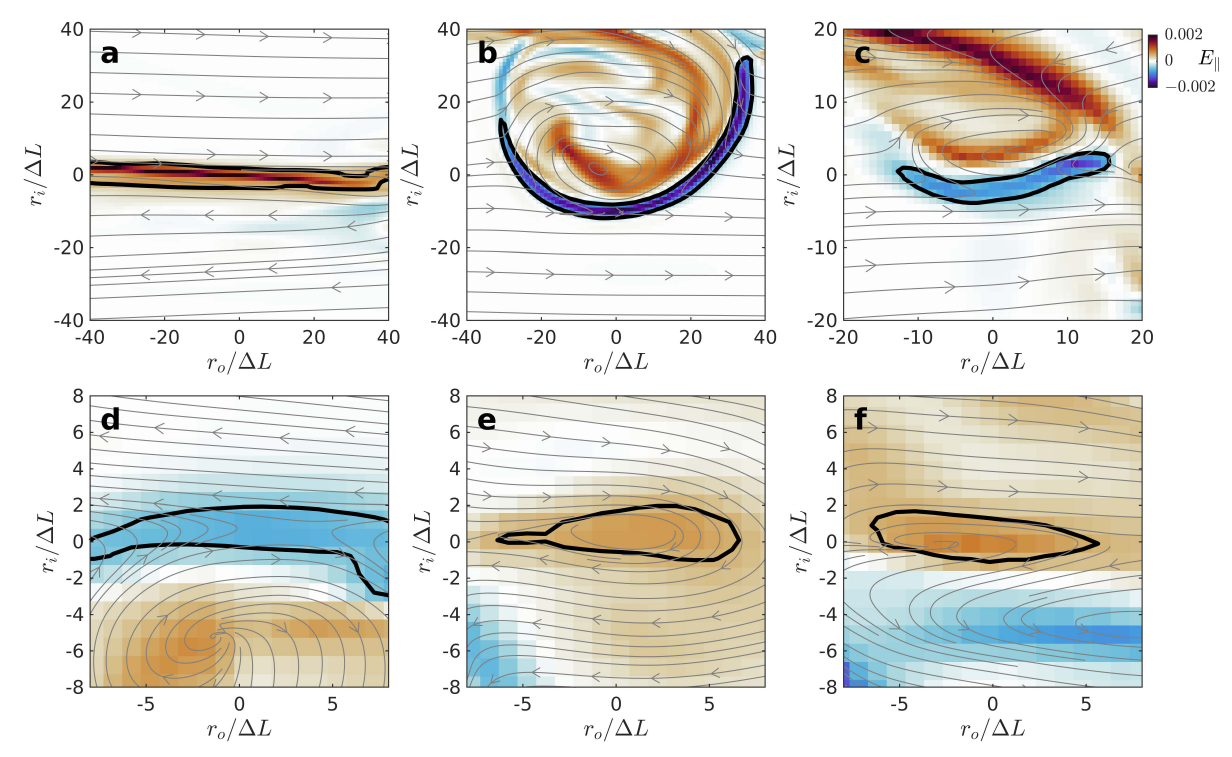}
\caption{Adjacent $E_\parallel$ and magnetic structures of six reconnection kernels in their local reconnection reference frames.
All slices pass the origin points of local reconnection frames, and $r_i$ and $r_o$ are coordinates along $\hat{\bf e}_i$ and $\hat{\bf e}_o$, respectively.
The thick black curves outline the inflow edges with $\lvert E_\parallel \rvert=E_\mathrm{thres}$, and the thin gray curves depict the project magnetic field lines.
\label{sfig_MRPSlice}}
\end{figure}

Using this reference frame, the reconnection properties of individual reconnection kernels can be effectively analyzed.
In Fig.\,\ref{sfig_MRPSlice}, we present the profiles of $E_\parallel$ and the magnetic field lines for six reconnection kernels on the $\hat{\bf e}_i$-$\hat{\bf e}_o$ planes, analogous to the slice shown in Fig.\,\ref{fig:MRP_Diagram}c.
Panels a--c correspond to three relatively large reconnection kernels, while panels d--f depict smaller ones.
Despite variations in the surrounding magnetic structures and spatial scales, all reconnection kernels are properly aligned within their respective local reference frames, facilitating a systematic and simplified analysis of their geometric characteristics and reconnection properties.

\section{Statistical properties at different moments}\label{apdx:difftime} %
\begin{figure}[htb]
\centering
\includegraphics[width=1\textwidth]{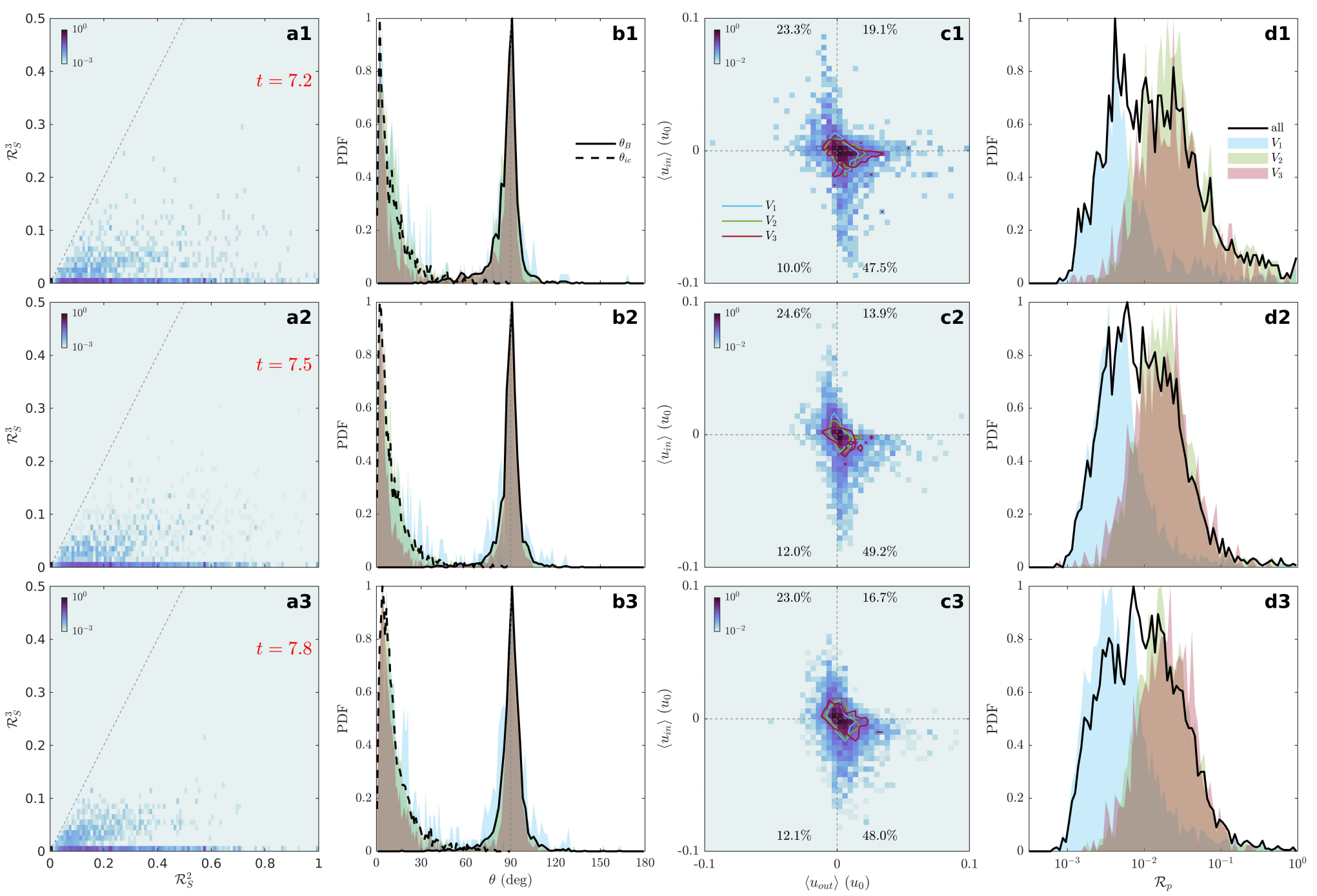}
\caption{Statistical properties of reconnection kernels at different moments.
The first, second, and third rows show results at $t=7.2$, $7.5$, and $7.8$, respectively.
The first column shows the joint PDFs of $\mathcal{R}_S^2$ and $\mathcal{R}_S^3$ for reconnection kernels with finite $L_1$.
The second column presents the PDFs of $\theta_B$ (right peaks) and $\theta_{ic}$ (left peaks) for reconnection kernels with finite $L_2$.
The third column displays the joint PDFs of $\langle u_{out}\rangle$ and $\langle u_{in}\rangle$.
The fourth column illustrates the PDFs of $\mathcal{R}_p$.
The colored curves and shades have the same meaning as in Figs.\,\ref{fig:MRP_Shape} and \ref{fig:MRP_V_Xi}.
\label{sfig_diffTime_shapeXi}}
\end{figure}

In the main text, we focus on the final moment at $t=8.2$ to analyze the statistical properties of a well-developed, strongly turbulent state.
Figure \ref{sfig_diffTime_shapeXi} presents the statistical results at three earlier moments.
At these earlier stages, the magnitude of perturbation magnetic fields is lower, and the guide field strength in the $z$-direction is relatively stronger.
Nevertheless, despite the varying nature of reconnecting CS toward the fully turbulent state, the statistical properties of reconnection kernels remain remarkably consistent with that at $t=8.2$.

\section{Influences of $E_\mathrm{thres}$}\label{apdx:diffE} %
Here we examine the impact of different threshold $E_\mathrm{thres}$ on the statistical results.
Setting $E_\mathrm{thres}$ to zero results in prohibitively high computational costs and excessively large sizes of reconnection kernels, complicating the analysis, particularly for field-line tracing.
Conversely, if $E_\mathrm{thres}$ is too large, on the one hand, reconnection kernels become excessively fragmented, leading to a reduction in sample size that may not adequately reflect the statistical properties.
On the other hand, many regions with finite $E_\parallel$ will be missed, which causes the underestimation of the total reconnection rate.
Therefore, our approach for selecting $E_\mathrm{thres}$ adheres to two key principles.
First, we aim to keep computational costs at an acceptable level.
Second, we ensure that the statistical results remain robust against reasonable variations in $E_\mathrm{thres}$.
To validate the robustness of our findings, we have tested two additional threshold values $E_\mathrm{thres}=2.5\times 10^{-4}$ and $7.5\times 10^{-4}$.

\begin{figure}[htb]
\centering
\includegraphics[width=1\textwidth]{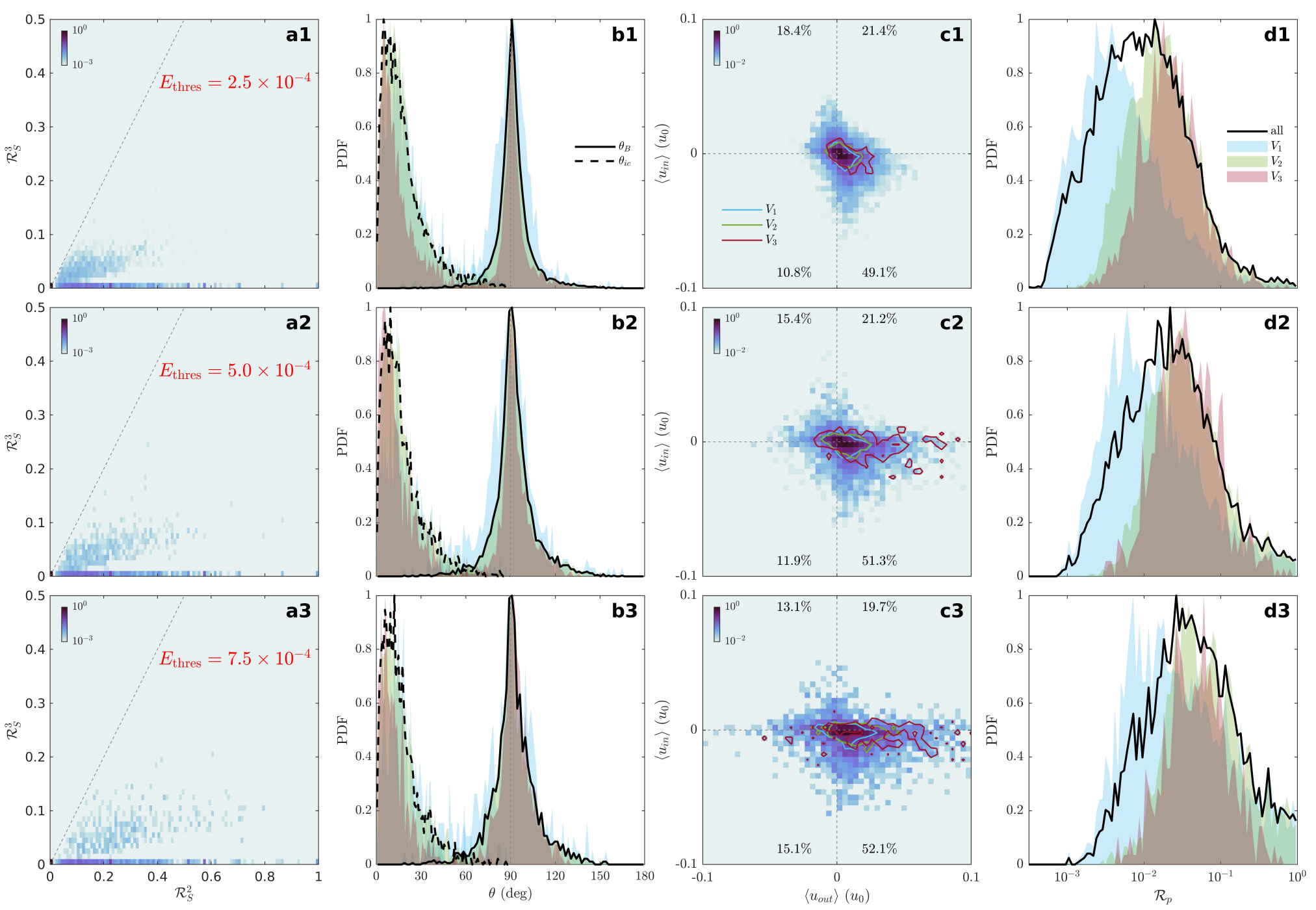}
\caption{Statistical properties of reconnection kernels for different $E_\mathrm{thres}$ values at $t=8.2$.
The first, second, and third rows correspond to results obtained with $E_\mathrm{thres}=2.5\times 10^{-4}$, $5.0\times 10^{-4}$ (used in the main text), and $7.5\times 10^{-4}$, respectively.
The arrangement of the columns follows the same order as in Fig.\,\ref{sfig_diffTime_shapeXi}.
\label{sfig_diffEthres_shapeXi}}
\end{figure}

As illustrated in Fig.\,\ref{sfig_diffEthres_shapeXi}, the statistical properties of shapes (the first two columns), reconnection flows (the third column), and reconnection rates (the final column) exhibit only minor variations with changes in $E_\mathrm{thres}$.
Meanwhile, for all cases of $E_\mathrm{thres}$, the statistical results are approximately independent of the sizes of reconnection kernels.
That is to say, the results presented in the main text are robust to variations in $E_\mathrm{thres}$.

\begin{figure}[htb]
\centering
\includegraphics[width=0.9\textwidth]{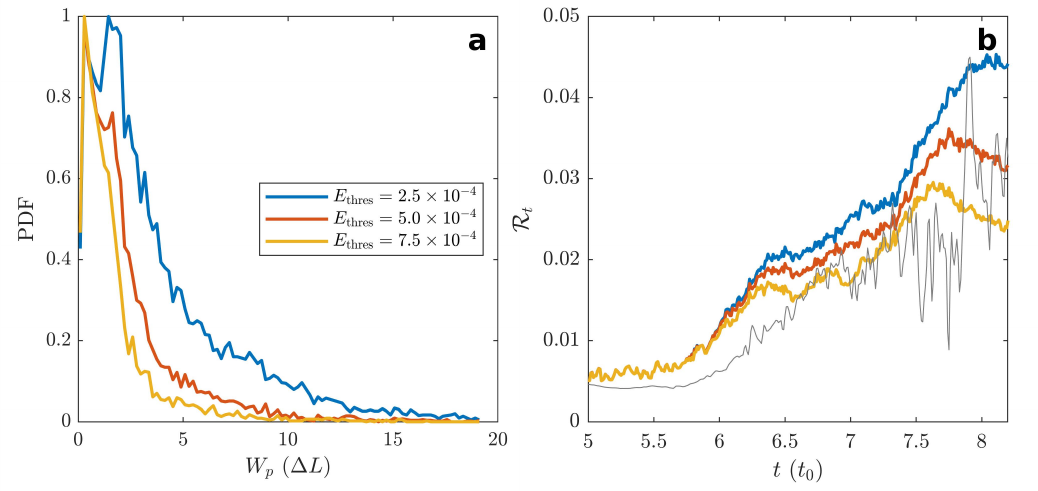}
\caption{PDFs of $W_p$ (a) and temporal evolutions of $\mathcal{R}_t$ (b) obtained by different $E_\mathrm{thres}$.
Colored curves correspond to different cases of $E_\mathrm{thres}$.
The gray curve in panel (b) depicts the evolution of $\mathcal{R}_{\bar{\bf B}}$.
\label{sfig_diffEthres_WpXiall}}
\end{figure}

There are two primary effects of $E_\mathrm{thres}$.
First, a smaller $E_\mathrm{thres}$ results in a larger average width sandwiched by inflow edges (see Fig.\,\ref{sfig_diffEthres_WpXiall}a).
Second, a smaller $E_\mathrm{thres}$ allows for the inclusion of more regions with weak reconnection, and also increases the volume of individual reconnection kernels, thereby extending the integration paths of $\Xi_\mathrm{max}$ lines.
As a result, $\mathcal{R}_t$ increases with decreasing $E_\mathrm{thres}$ despite not significantly (see Fig.\,\ref{sfig_diffEthres_WpXiall}b).
It should be noted that, for the smallest $E_\mathrm{thres}$ case, $\mathcal{R}_t$ exhibits a prolonged rising phase after $t=7.5$, eventually reaching a plateau at $t=8$.
In contrast, for the other two cases, $\mathcal{R}_t$ starts to decline after approximately $t=7.8$. This behavior primarily arises from two factors.
On the one hand, the turbulence further produces numerous smaller-scale reconnection fragments with weaker reconnection rates (as shown by the blue shade in Fig.\,\ref{sfig_diffEthres_shapeXi}d1--d3). Including these smaller fragments increases the total reconnection rate.
On the other hand, as these smaller-scale fragments approach the resolution limit of the simulation, numerical errors in the analysis become increasingly significant, which might cause a larger uncertainty of the reconnection rate.
To obtain a more accurate reconnection rate during the later stages, higher-resolution simulations are required.
However, the overall trends in the PDF of $W_p$ and the evolution of $\mathcal{R}_t$ are not significantly unaffected.

\section{Fractal Analysis of Reconnection Kernels}\label{apdx:fd} %

The procedure of fractal analysis is similar with \cite{Isliker2019}.
To perform the box-counting analysis, we first assigning a value of one to $E_\mathrm{max}$ grids and zero to all other grids within the CS region defined by $x\in\left[-0.05, 0.05\right]$, $y\in\left[0.45, 1\right]$, and $z\in\left[-0.15, 0.15\right]$.
The resulting dataset is then analyzed using the open-source box-counting script developed by \cite{Moisy2008}, which outputs the relationship between box size $L_b$ and the number of covering boxes $N_b$, as shown in Fig.\,\ref{fig:DF}.
During the analysis of fractal dimension, we exclude the data of box sizes larger than the CS width to avoid the influences of their numerical errors.

\begin{figure}[htb]
\centering
\includegraphics[width=0.9\textwidth]{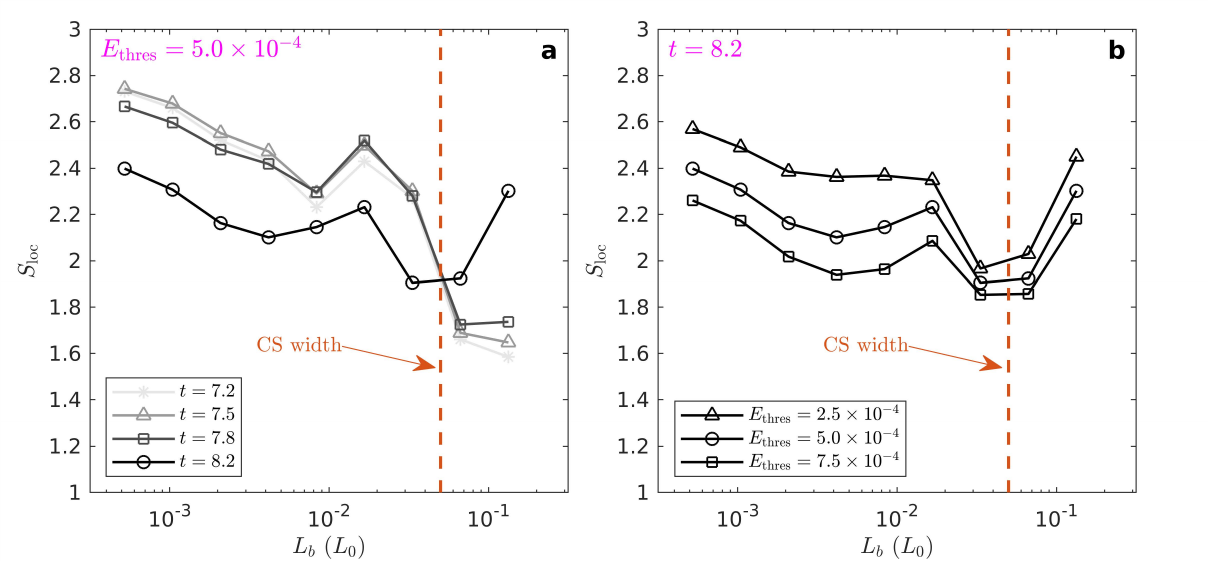}
\caption{Influences of time and $E_\mathrm{thres}$ on the results of box-counting analysis.
(a) Local scaling exponents obtained at $t=7.2$, $7.5$, $7.8$, and $8.2$ for $E_\mathrm{thres}=5.0\times 10^{-4}$.
(b) Local scaling exponents calculated at $t=8.2$ for $E_\mathrm{thres}=2.5\times 10^{-4}$, $5.0\times 10^{-4}$, and $7.5\times 10^{-4}$.
The orange dashed lines denote the maximum $z$-direction width of the CS region at $t=8.2$.
\label{sfig_DF_Diff_t_Diff_Ethres}}
\end{figure}

In Fig.\,\ref{sfig_DF_Diff_t_Diff_Ethres}, we compute the local scaling exponent defined by $S_\mathrm{loc}=-\mathrm{d}lnN_b/\mathrm{d}lnL_b$, and study the influences of time and $E_\mathrm{thres}$ on the results of fractal analysis.
At $t=8.2$, the $S_\mathrm{loc}$ curve exhibits an approximately constant plateau extending over nearly one order of magnitude around $L_b=10^{-2}$, confirming the emergence of fractal scaling behavior following the onset of turbulence (see the black curve with circle markers in Fig.\,\ref{sfig_DF_Diff_t_Diff_Ethres}a). 
At earlier times, $S_\mathrm{loc}$ displays greater fluctuations compared to the later stage; however, it still suggests the presence of approximate fractal characteristics.  
Prior to the development of turbulence, the reconnection region remains relatively unfragmented, leading to larger filling factors than those observed at $t=8.2$ (Fig.\,\ref{sfig_DF_Diff_t_Diff_Ethres}a).
Moreover, varying the threshold value $E_\mathrm{thres}$ has a negligible impact on the global trend of $S_\mathrm{loc}$ and primarily affects its absolute value.
Specifically, a smaller $E_\mathrm{thres}$ incorporates more $E_\mathrm{max}$ grids in the 3D CS region, thereby increasing both the filling factor and $S_\mathrm{loc}$ (Fig.\,\ref{sfig_DF_Diff_t_Diff_Ethres}).

\section*{Acknowledgments}
This research is supported by the Strategic Priority Research Program of the Chinese Academy of Sciences (Grant No. XDB0560000), the Natural Science Foundation of China (Grant No. 12473057 and 12127901), and the National Key R\&D Program of China (Grant No. 2021YFA1600504).
The simulation is performed at the National Supercomputing Center in Jinan and the data analysis is supported by the High Performance Computing Center (HPCC) of Nanjing University.

\bibliography{library}
\bibliographystyle{aasjournal}
\end{document}